# A portable electronic system for in-situ measurements of oil concentration in MetalWorking fluids

M. Grossi[*], B. Riccò

[*] Corresponding author : marco.grossi8@unibo.it , Tel. 0039-0512093078 , Fax 0039-0512093785

Department of Electrical Energy and Information Engineering "Guglielmo Marconi" (DEI), University of Bologna, Bologna, Italy

**Abstract**

MetalWorking Fluids (MWFs) are widely used to cool and lubricate machines and tools. By far, the most common MWFs are oil-in-water emulsions with oil concentration ($C_{oil}$) in the range from 1% to 10% , depending on type of oil, material to be worked, etc. In order to optimize emulsion and machine performance, as well as for good waste policy, the right value of $C_{oil}$ should be kept (approximately) constant during the MWF's lifecycle to compensate inevitable changes due to water evaporation, bacterial attack, oil adhesion to metal parts, etc.. This, however, requires periodic measurements, often skipped because they require unhandy operations and produce inaccurate results. In this context, a new system is presented that is based on the falling ball principle, normally used for viscosity measurements, shown to be suitable also for accurate $C_{oil}$ measurements. In our system, the transit time of the sphere within the instrument is determined by means of inductive proximity sensors, a PT100 sensor is used for temperature whose effects on $C_{oil}$ are accounted for by means of an ad-hoc algorithm. A battery-operated electronic board has been designed that allows rapid and user-friendly "in-situ" measurements and a prototype of an whole portable and automatic instrument, suitable for in-situ measurements, has been fabricated.

## 1. Introduction

MetalWorking Fluids (MWFs), widely used in industries for machine cooling and lubrication during different finishing processes [1], usually are mix of water and soluble oils with oil concentration ($C_{oil}$) in the range from 1% to 10%, depending on oil type, material to be worked etc. In addition, they often contain additives, such as anti-microbial compounds (to prevent bacterial degradation) [2][3], corrosion inhibitors, emulsifier, pressure additives and anti-foam agents.

The oil concentration should be optimized in order to prevent potential malfunctions and reliability problems. In particular, if $C_{oil}$ is too low, poor lubrication can lead to corrosion of machines, tools and metal parts under work, as well as to proliferation of microflora; on the contrary, too high values of $C_{oil}$ can produce foaming, tool malfunctions due to excessive lubricity and potential health problems for the exposed workers (since oil compounds tend to be dispersed in the environment as aerosols) [4][5][6].

Unfortunately, under normal working conditions, $C_{oil}$ tends to change due to water evaporation, bacterial attack, oil adhesion to metal parts, etc. Thus, oil concentration should be monitored at regular intervals to compensate excessive deviations from its optimal value.

The official technique to measure $C_{oil}$ in MWF samples is (manual) titration [7]: 100ml of the fluid under test is titrated with a 0.5M HCl solution to an endpoint of pH= 4 and the volume of titrant necessary to reach such an endpoint is used to determine $C_{oil}$. This technique is accurate and substantially unaffected by the contaminations inevitably occurring when the MWFs are used, but must be performed by trained personnel in a laboratory: hence, it is not suitable for "in-situ" measurements within industrial plants.

For this reason, the standard technique used in industry is refractometry [8]: a few drops of MWF are placed on the refractometer prism and exposed to light; the fluid refractive index is then read on the instrument scale and converted into oil concentration by means of the Brix scale (quantifying the solid content dissolved in the emulsion). Refractometers can be used within production environments and are also relatively cheap. However, their most common manual versions (costing about 200 US$) present a few drawbacks, since: they need to be frequently calibrated; the accuracy

depends on the user's skill and is affected by fluid contamination; samples with certain composition do not provide clear readings; automatic compensation for the effect of temperature (T) is not provided.

Recently, portable digital refractometers (with a cost of 300 to 400US$), featuring temperature compensation and digital display, have been introduced into the market. While these instruments address some of the problems of analog refractometers, when dealing with (long) used MWFs accuracy is still a problem because of the influence of solid contaminants on the Brix scale. Furthermore, working with a few drops of fluid, refractometers have problems of statistical significance of the results.

On the other hand, low-cost portable instruments for accurate "in-situ" measurements are desirable for quality control in industrial productions and a number of solutions have been proposed in different fields, such as, for instance: gas chromatography for analysis of combustible gases in coal mining [9]; multi-sensor data-logging for medical surveillance in harsh environments [10]; automatic characterization of ice-cream mixes by impedance spectroscopy [11]; electronic system to control ice-cream freezing by electrical characteristic analysis [12]; biosensors for glucose measurements [13]; electronic sensor systems for acidity [14][15], peroxide value and total phenolic content [16] in olive oil.

In this context, this work presents a novel instrument to measure oil concentration in MWFs that exploits the falling ball principle, mainly used for fluid's viscosity but never before applied to the case of interest here. To this purpose, the time ($t_{fall}$) required for a sphere of known density and volume to fall under gravity for a given distance within the sample of interest is automatically measured and converted into an accurate value of $C_{oil}$.

To reach this goal, first the applicability of the falling ball principle to the case of interest has been studied by means of a bench-top set-up then, based on the positive results of the experiments, a fully automatic, battery-operated portable instrument has been developed that can be used by anybody and in any environment for quick and accurate "in-situ" measurements of $C_{oil}$ in MWFs.

Such an instrument presents substantial innovations with respect to most commercially available falling ball viscometers, in that: 1) the sphere transit time is automatically detected by means of inductive proximity sensors; 2) measurements are carried out at room temperature (T), instead of at a standard temperature. This requires that T be measured (in our case, by means of the integrated PT100 sensor) so as to account for the effects of temperature (by means of a suitable algorithm). Avoiding a (controlled) heating system, this solution reduces the power consumption making it suitable for a battery-operated, portable instrument.

Beside the application addressed in this paper, the instrument presented in this paper has many other possible applications, since liquid viscosity is an important parameter in various industrial fields, such as, for instance: food quality assurance [17]; cosmetics [18]; paints [19]; oil [20].

The rest of the paper is organized as follows. Section 2 illustrates the application of the falling ball method to the measurement of $C_{oil}$ in MWFs. Section 3 describes the experiments done with a bench-top set-up in order to validate the method introduced in the previous Section. Section 4 describes the portable instrument developed for the application of interest here. Finally, conclusions are drawn in Section 5.

## 2. Materials and methods

*2.1. The falling ball approach for oil concentration measurement in MWFs*

Viscosity is an important property of fluids that can be measured with a variety of scientific instruments [21] among which the falling ball principle of interest for this work.

A sphere moving in a fluid is subject to three forces, as depicted in Fig. 1 (a) [22], namely: gravity ($F_G$), buoyancy ($F_B$) and viscosity ($F_V$).

$F_G$ (acting downward) is the product of the sphere mass and the gravitational acceleration:

$$F_G = m_s \cdot g = \frac{4}{3} \pi \cdot r^3 \cdot \rho_2 \cdot g \quad ; \tag{1}$$

$F_B$, exerted by the fluid and opposing the sphere's motion, can be expressed as:

$$F_B = m_l \cdot g = \frac{4}{3}\pi \cdot r^3 \cdot \rho_1 \cdot g \quad ; \tag{2}$$

finally, $F_V$, also opposing the sphere's motion, can be expressed as:

$$F_V = 6\pi \cdot \eta \cdot r \cdot v \quad . \tag{3}$$

In the above Equations: $m_s$ is the sphere mass; $m_l$ the mass of the liquid displaced by the ball; $\rho_2$ the sphere density; $\rho_1$ the liquid density; r the sphere radius; g the gravity acceleration; η the dynamic viscosity of the liquid; v the sphere velocity.

Since, after a short transient (negligible in our case), the sphere falls at constant speed, it must be:

$$F_G = F_B + F_V \quad , \tag{4}$$

then:

$$\frac{4}{3}\pi \cdot r^3 \cdot \rho_2 \cdot g = \frac{4}{3}\pi \cdot r^3 \cdot \rho_1 \cdot g + 6\pi \cdot \eta \cdot r \cdot v \quad . \tag{5}$$

Denoting with L and $t_{fall}$ the length and the duration of the sphere fall, respectively, it is:

$$\eta = \frac{2}{9} \cdot \frac{r^2}{L} \cdot g \cdot (\rho_2 - \rho_1) \cdot \mathbf{t_{fall}} \quad , \tag{6}$$

thus:

$$\eta = k \cdot (\rho_2 - \rho_1) \cdot \mathbf{t_{fall}} \quad , \tag{7}$$

where k is a parameter depending only on geometrical parameters.

The model of Eq. (7) has been derived for a sphere falling in an infinite volume of liquid. However, it also holds if it falls within a cylinder, since the wall-effects can be accounted using the Ladenburg correction to adjust the k parameter, that becomes dependent also on the ratio between the cylinder and ball diameter [22].

In turn, as discussed in more detail in Section 3.1, the sample viscosity η is found to be a linear function of $C_{oil}$, hence:

$$C_{oil} = \alpha \cdot t_{fall} + \beta \quad , \tag{8}$$

where α and β are empirical parameters (depending on instrument geometry, type of products, temperature etc.), in our case to be determined by means of a calibration process (as discussed in Section 4).

*2.2. Experimental set-up*

The set-up used for the experiments, shown in Fig. 1 (b) features an Aluminum block with a cylindrical cavity of radius R = 7.265 mm, chosen after preliminary tests on different values, to be filled with the MWF under test. A chrome steel, high precision ball (of radius r = 7.2 mm) is allowed to fall into the cavity and the time ($t_{fall}$) taken to cover the distance between two points (distant 150 mm from one another) is measured by means of two inductive proximity sensors (Fig. 1 (c)), powered by a single 5V DC power supply and providing a digital output: $V_{out}$ = 5V or 0 if the sphere is present or absent at near distance (less than about 2 mm), respectively.

A PT100 thermistor, with the probe tip touching the MWF inside the cavity, is used to measure the liquid temperature (T), necessary to account for viscosity variations due to temperature.

A support has been created to allow different inclination of the cylindrical cavity with respect to the vertical standing.

The experimental set-up used for the experiments is shown in Fig. 2. A PC-based system with the data acquisition board NI USB-6211 (by National Instruments) acquires the digital outputs of the top and bottom proximity sensor ($V_{OUT,TOP}$ and $V_{OUT,BOTTOM}$, respectively), as well as the analog output of the temperature sensor ($V_{OUT,TEMP}$). The DC power supply Agilent E3631A provides the +5V supply for the proximity sensors and the ± 10V dual supply for the temperature measuring circuit.

The temperature is measured by means of a Wheatstone bridge and a differential amplifier. The PT100 sensor is a thermistor whose resistance is linearly related to the temperature according to the formula $R_{PT100}$ = 0.3844·T + 104.3, where $R_{PT100}$ is expressed in Ω and T in °C. The PT100 thermistor is inserted in one of the four branches of the bridge as shown in Fig. 2. Since $R_{10k\Omega}$ >> $R_{100\Omega}$ the voltage difference $V_2 - V_1$ can be estimated as:

$$V_2 - V_1 \approx \frac{10}{R_{10k\Omega}} \cdot (R_{PT100} - R_{100\Omega}) \qquad (9)$$

The output of the differential amplifier can expressed as:

$$V_{OUT,TEMP} = \left(1 + \frac{2R_1}{R_{GAIN}}\right) \cdot \frac{R_3}{R_2} \cdot (V_2 - V_1) \qquad (10)$$

Using the calibration Equation for PT100, Eqs. (9) and (10) allow to calculate T as function of $V_{OUT,TEMP}$.

All software for data acquisition, display and filing has been developed using the programming language LabVIEW (National Instruments).

A thermal incubator, Binder APT KB 53, has been used for tests at different temperatures.

## 3. Results and discussion

For the experiments of this work samples of MWFs featuring oil concentration in the range 1% − 10% have been used. Each sample has been tested in triplicate and mean value as well as standard deviation of the data have been calculated. When not otherwise specified, the samples were created using the oil product "Spirit MS 8200" by Total.

### 3.1. Dependence of $t_{fall}$ vs. $C_{oil}$

Fig. 3 (a) shows $t_{fall}$ vs. $C_{oil}$ for different samples, with the cylinder inclined at an angle of 10° with respect to the vertical standing and T = 25 °C. As can be seen, a good linear relationship is found between $t_{fall}$ and $C_{oil}$, that can be expressed with the regression line equation $t_{fall} = 0.5946 \cdot C_{oil} + 19.967$.

Thus, Eq. 8 can be written as:

$$C_{oil} = 1.682 \cdot t_{fall} - 33.58, \qquad (11)$$

with a coefficient of determination $R^2 = 0.989$.

Such a relationship means that, for $C_{oil}$ ranging from 0% to 10%, $t_{fall}$ varies from 19.967 to 25.913 seconds (corresponding to a 29.78% variation).

Since, as from Eq. 7, $t_{fall}$ depends on both $\eta$ and $\rho_2 - \rho_1$, MWF samples featuring different oil concentration have been measured to test the sensitivity of $\eta$ and $\rho_1$ on $C_{oil}$. These experiments have been carried out as follows: the weight of an empty precision flask has been measured using an analytical balance, then the flask has been filled with 100 cm$^3$ of the sample under test and its weight measured again. Thus the weight of the sample is given by the difference between the two results, then the sample density $\rho_1$ is easily calculated.

The measurements have been carried out at room temperature (25 °C) and the results are shown in Figure 3 (b). As can be seen, the sample density is also a linear function of $C_{oil}$, although the slope of the regression line is much smaller than in the case $t_{fall}$ vs. $C_{oil}$. The sample density $\rho_1$ varies from 0.9926 g·cm$^{-3}$ to 0.987 g·cm$^{-3}$ as $C_{oil}$ ranges from 0% to 10%. Given for $\rho_2$ a value of 7.85 g·cm$^{-3}$ (the chrome steel density), the difference $\rho_2 - \rho_1$ varies of only 0.08116% as $C_{oil}$ ranges from 0% to 10%. This clearly indicates that almost the whole variation of $t_{fall}$ is due to the variation of the fuid viscosity, while the contribution of the density is almost negligible.

*3.2. Effects of inclination angle*

A set of experiments has been carried out to test the effects of the cylinder inclination. Denoting with $\vartheta$ the angle between vertical and cylinder axis (Fig. 4 a), five values of $\vartheta$ have been tested: 0° (instrument in the vertical position), 10°, 25°, 55° and 65°. For each inclination the measurements have been carried out on the same MWFs set of Fig. 3 (a), featuring: $C_{oil}$ = 1.96%, 4.46%, 7.12% and 10.17% ; T = 25°C.

The results of Fig. 4 (b) show a very good linear correlation ($R^2 > 0.987$) between $t_{fall}$ and $C_{oil}$ with regression slope increasing with $\vartheta > 0$ (0.5946 for 10°, 0.7 for 25°, 1.1378 for 55°, 1.609 for 65°). Fig. 4 (c) shows $t_{fall}$ vs. $1/\cos(\vartheta)$ for the sample featuring $C_{oil}$ = 1.96% and a good linear correlation is obtained: y = 18.335·x + 2.1217, $R^2$ = 0.9972. In this situation the sphere rolls on the instrument internal surface (as depicted in Fig. 4 (a)). Thus; denoting with F the absolute value of the force acting on the sphere, and with $F_{\parallel} = F \cdot \cos(\vartheta)$ the force component in the direction of motion, it is:

F· cos($\vartheta$) = $m_s$·$a_\parallel$ = $m_s$·$dv_\parallel/dt$, hence the proportionality between $t_{fall}$ and 1/cos($\vartheta$), where $a\parallel$ and $v\parallel$ are the components of acceleration and velocity in the direction of motion.

Instead, when $\vartheta$ = 0° the situation is different because the sphere does not roll but falls perpendicularly, randomly bumping on the cavity surface. This case is depicted in Fig. 4 (d), where R − r is the average distance between the sphere and each side of the instrument surface, while the case $\vartheta$ > 0° is sketched in Fig. 4 (e).

As a consequence of the bumpy motion, for the same $C_{oil}$, $t_{fall}$ for $\vartheta$ = 0° is higher and also less repeatable (± 450 ms vs. ± 250 ms) than in the case $\vartheta$ = 10°.

Higher values of $\vartheta$ result in higher sensitivity, but the sphere occasionally gets stuck in the cavity, especially when old and contaminated emulsions are tested. Thus, $\vartheta$ = 10° has been chosen as optimal condition.

*3.3. Effects of temperature*

Since liquid viscosity is strongly dependent on temperature, experiments have been done at T= 15°C, 20°C, 25°C, 30°C and 35°C (with $\vartheta$ = 10°) to study the effects of T on $t_{fall}$.

The results presented in Fig. 5 (a) show a very good linear correlation between $t_{fall}$ and $C_{oil}$ ($R^2$ > 0.98) with the following regression lines: at 15°C y = 0.8657·x+24.79; at 20°C y = 0.6636·x+22.59; at 25°C y = 0.5946·x+19.96; at 30°C y = 0.4993·x+17.97; at 35°C y = 0.4725·x+16.37. As can be seen the sensitivity increases as T decreases.

A mathematical model has then been developed to describe the variation of $t_{fall}$ with both $C_{oil}$ and T. The simplest way to describe the variation of viscosity with temperature, over a limited range of values, is the empirical model proposed by Reynolds in 1886 [24]:

$$\eta_T = \eta_0 e^{-b \cdot (T-T_0)}, \qquad (12)$$

where $\eta_0$ is the viscosity at temperature $T_0$ and b is an empirical parameter.

Since, according to Eq.(7), the viscosity is linearly related to $t_{fall}$ and the density variation of oil concentration is negligible, we can write:

$$t_{fall,T} = t_{fall,T0} e^{-b \cdot (T-T_0)} , \qquad (13)$$

where $t_{fall,T0}$ is the measured falling time at the temperature $T_0$.

Fig. 5 (b) show $t_{fall}$ vs. T for each of the four samples tested in this work and the data have been fitted using Eq. (13) obtaining very good correlations ($R^2 > 0.99$). Our results also show that the empirical parameter b of Eq. (13) increases almost linearly ($R^2 = 0.847$) with $C_{oil}$ according with the Equation:

$$b = 166.363 \cdot 10^{-6} \cdot C_{oil} + 21.212 \cdot 10^{-3} . \qquad (14)$$

*3.4. Calibration for different products*

Finally, measurements have been carried out to test the set-up response to MWF samples created with different types of oil. Fig. 6 shows $t_{fall}$ vs $C_{oil}$ for two different products ("Spirit MS 8200" by Total and "Adrana D 208" by Houghton) at T = 25°C and $\vartheta$ = 10°. As can be seen the calibration line is different for the two products, in particular "Spirit MS 8200" is more sensitive than "Adrana D 208" due to its higher viscosity.

**4. The automatic and portable instrument**

Based on the results of Section 3, a portable electronic instrument has been designed for quick, in-situ measurements of MWFs oil concentration.

All the measurement functions have been realized by means of an ad-hoc electronic board based on the microcontroller STM32F103R6 (featuring a clock frequency of 16MHz). In particular, the proximity sensors activation is detected by means of two digital inputs ($D_0$ and $D_1$) of the microcontroller and T is measured acquiring the corresponding sensor voltage with an analog input (12 bits ADC). The system, operated with four 1.5V AAA alkaline batteries, features a monochromatic 2 lines x 8 columns text LCD and four buttons for user interaction. The mean current drawn by the instrument during operation is 28 mA (with LCD backlight disabled). Pictures of the instrument, costing about 180US$ , are shown in Fig. 7 (a) while the essential schematics of the electronic part is shown in Fig. 7 (b).

Before operation, the instrument must be calibrated for the particular product of interest. To this purpose, three MWF emulsions featuring $C_{oil}$ of 2%, 6% and 10%, respectively are used and $t_{fall}$ for these cases is measured. Then, Eq. 8 is used as the calibration line and the two parameters α and β are calculated. These parameters as well as the calibration temperature $T_{calib}$ are stored on the board $E^2PROM$.

When in use, the instrument must be manually filled with the MWF of interest and the configuration file of the tested product must be selected. Then $t_{fall}$ as well as T are measured and $t_{fall}$ is corrected for the effects of temperature by means of the model described in Section 3.3. Since, the temperature coefficient b is also function of $C_{oil}$ (Eq. 14), temperature compensation requires the iterative procedure presented in Fig. 7 (c): a guess value ($C_{oil}^*$) of oil concentration is initially set at 5 % and the b parameter is calculated with Eq. 14; then $t_{fall}$ is compensated for the difference between T and $T_{calib}$ using Eq. 13 and $C_{oil}$ is determined using the calibration line Equation. If $|C_{oil} - C_{oil}^*| < 0.1$ then $C_{oil}$ is considered as the final value, otherwise $C_{oil}^* = C_{oil}$ and the procedure is repeated. A video about use of the instrument is presented as Supplementary Video Data.

The designed instrument has been used to test a set of MWF samples, both fresh and used, with different levels of contamination and at different temperatures. The estimated values of $C_{oil}$ have been compared with the oil concentration of the samples, and the results presented in Fig. 8 indicate a satisfactory accuracy ($R^2 = 0.9695$).

A preliminary comparison between the instrument of this work and a refractometer (Reichert 10440 Automatic Temperature Compensated Hand Refractometer) has been done at environmental temperature (24 °C) using both fresh and used MWF samples, taken from actual industrial plants. Before any measurement, our instrument was calibrated as previously discussed, while the hand refractometer was used according to the user manual: initial calibration with water, cleaning with water between measurements. $C_{oil}$ is then obtained multiplying the measured value on the Brix scale with the refractometer index provided by the manufacturer.

Fresh MWF samples were created using two different products ("Spirit MS 8200" by Total and "Adrana D 208" by Houghton) and three different values of $C_{oil}$ ( 2%, 6% and 10%).

The values of $C_{oil}$ measured with both the compared instruments are shown in Figs. 9 (a) and 9 (b) in the cases of "Spirit MS 8200" by Total and "Adrana D 208" by Houghton, respectively.

In the former case, our instrument performs better for all concentrations as far as accuracy is concerned, while precision (i.e. differences in repeated measurements on the same sample) is better for concentrations $C_{oil}$ = 2% and 10% and comparable for $C_{oil}$ = 6%. The average relative error for repeated measurements was 0.0496 for our instrument and 0.0713 for the refractometer.

In the case of "Adrana D 208" the accuracy of our instrument is slightly lower for $C_{oil}$ = 2% and higher for $C_{oil}$ = 6% and 10%. As for precision, our instrument is comparable with the refractometer for $C_{oil}$ = 2% and 6%, while it is better for $C_{oil}$ = 10%. The average relative error for repeated measurements is 0.112 and 0.120 for our instrument and the refractometer, respectively.

Three "used" MWF samples (hereafter referred as sample #1, #2 and #3) have been used to compare the two instruments and a good linear correlation is found between the results obtained with them ($R^2$ = 0.9745). No comparison on accuracy could be done in the absence of "exact" values. As for precision, our instruments is better in the cases of samples #1 (0.0476 vs. 0.2) and #3 (0.0458 vs 0.199), while it is comparable for sample #2 (0.038 vs 0.035).

These tests indicate substantial agreement between the two instruments with significant advantages in terms of accuracy and precision for the one proposed in this work, expected to be superior than the other particularly in the case of used and worn-out MWFs since refractometry is sensitive to solid contaminants which do not substantially affect liquid viscosity. Moreover, since the volume of the SUT in our instrument is much larger than that in the refractometer (about 50 ml vs. a few drops), this results in better statistical significance of the results.

## 5. Conclusions

A battery-operated portable sensor to measure the oil concentration of MWFs has been developed. The instrument is based on the falling ball principle of viscosity measurement and estimates the oil concentration from the time needed by a high precision chrome steel ball to fall between two fixed points within a cavity filled with the product of interest. The sphere transit time is measured by means of a couple of inductive proximity sensors. A temperature sensor is integrated in the system to compensate for the effects of temperature.

The system has been tested with a set of MWF samples with oil concentration in the range 1% to 10% and the results show a good linear correlation between measured fall time and oil concentration.

The instrument of this work can be used for quick, "in-situ" measurements of oil concentration of MWFs, with benefits in terms of overall performance, reliability of both machines, tools and products, MWFs lifetime and waste management.

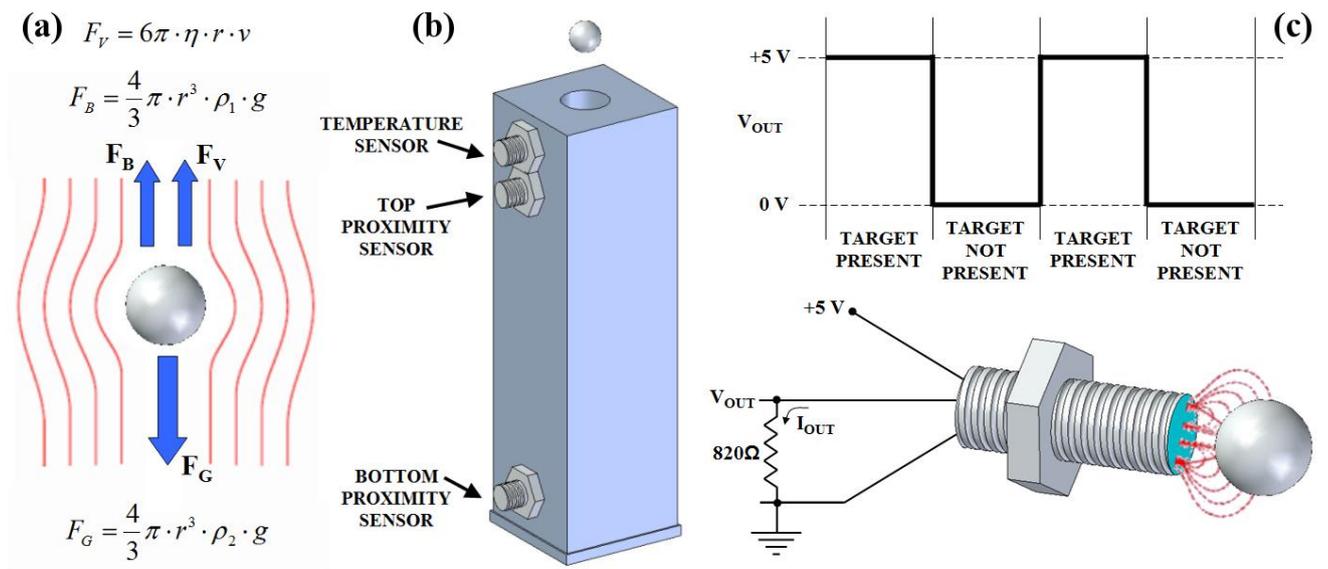

**Fig. 1**. (a) Forces acting on a sphere falling within a liquid medium, (b) a sketch of the designed instrument, (c) inductive proximity sensor used to detect the sphere transit.

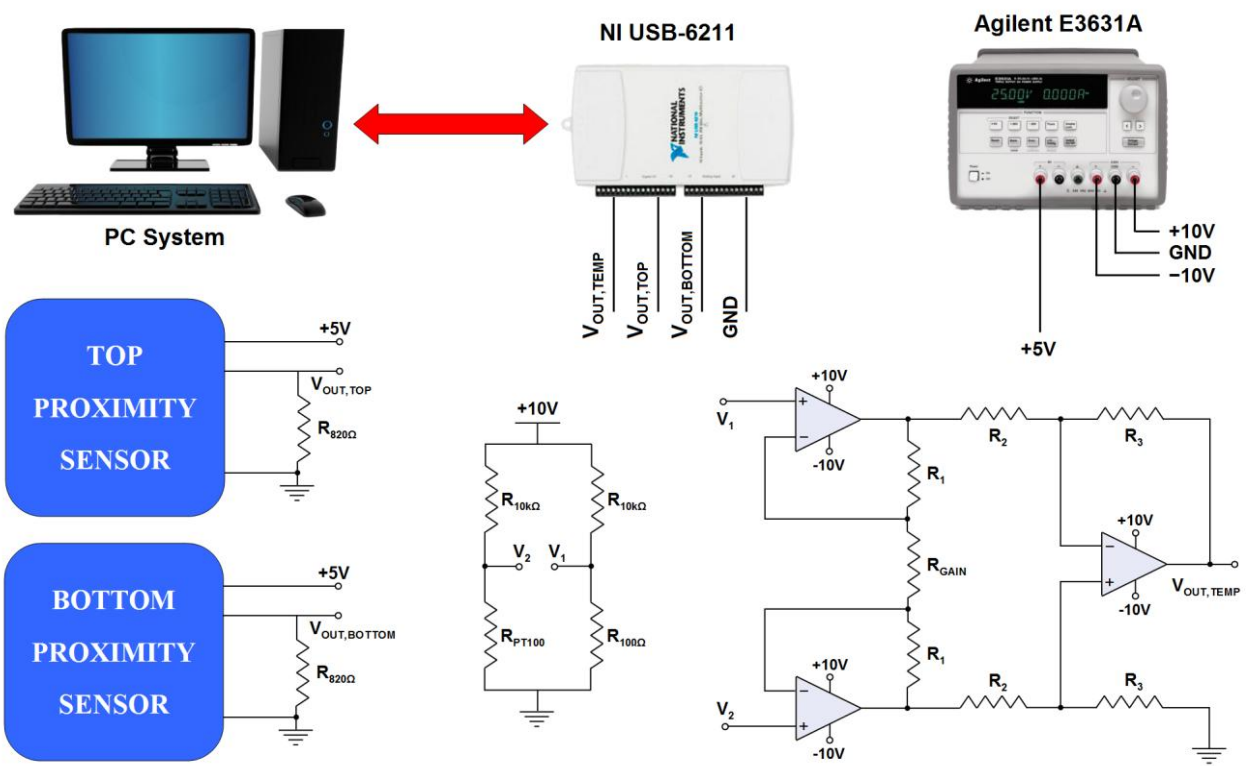

**Fig. 2**. Experimental set-up used to test the approach of this work.

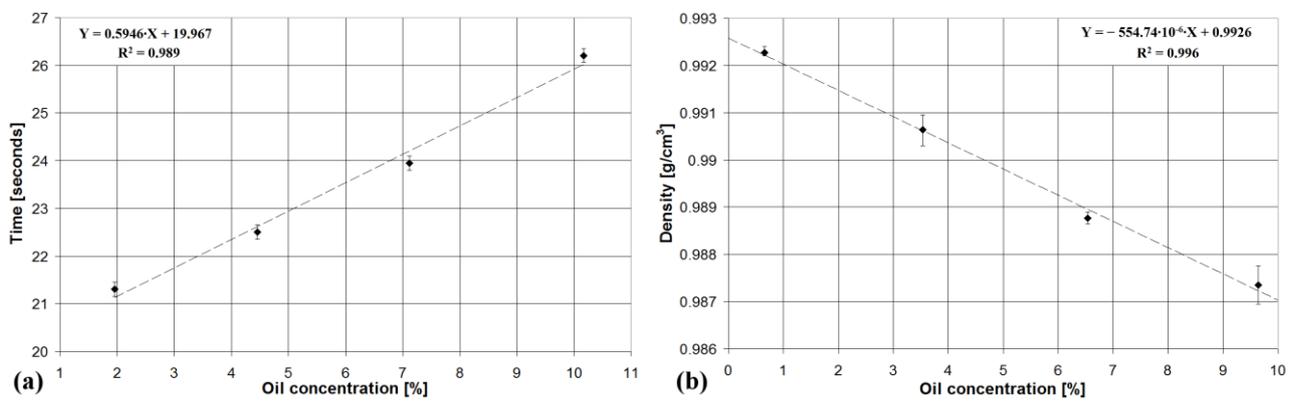

**Fig. 3**. (a) Measured $t_{fall}$ vs. $C_{oil}$ for an inclination angle of 10° and T = 25°C, (b) MWF sample density vs. $C_{oil}$ at T= 25°C.

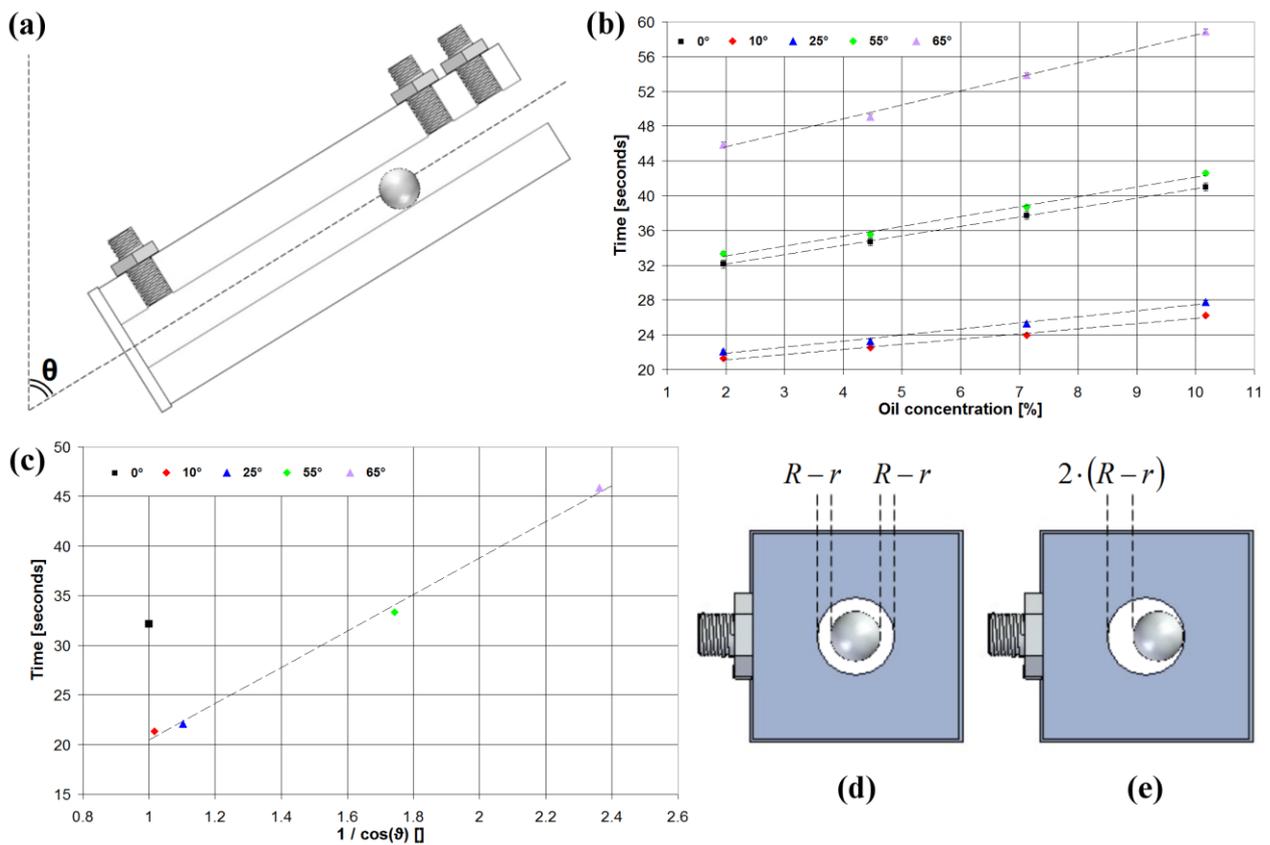

**Fig. 4**. (a) Side view of the instrument defining the inclination angle $\vartheta$, (b) measured $t_{fall}$ vs. $C_{oil}$ as function of $\vartheta$ for T = 25°C; (c) measured $t_{fall}$ vs. $1/\cos(\vartheta)$ for a sample with $C_{oil}$ = 1.96% and T = 25°C ; top view of the instrument in the case $\vartheta = 0$ (d) and $\vartheta > 0$ (e), respectively.

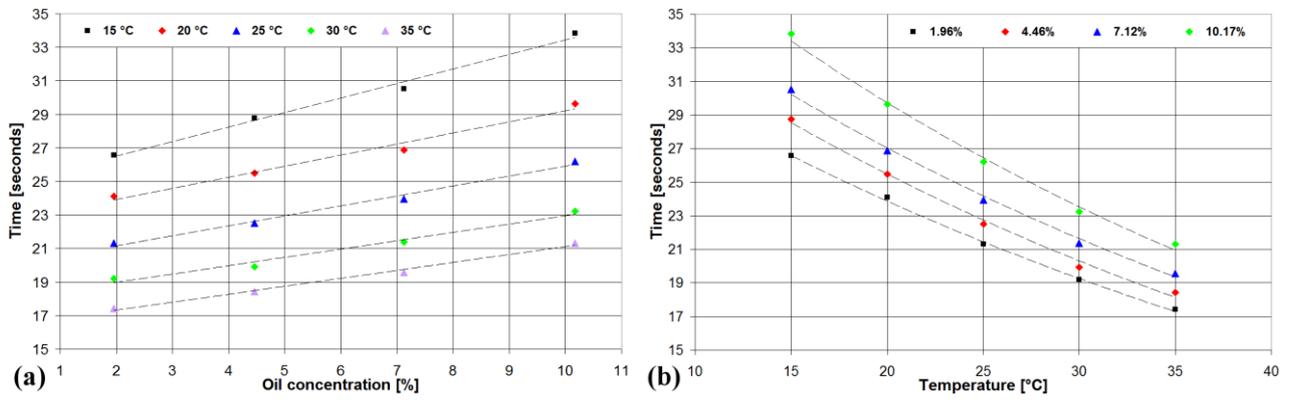

**Fig. 5**. (a) Measured $t_{fall}$ vs $C_{oil}$ for $\vartheta = 10°$ and different temperatures, (b) measured $t_{fall}$ vs T for different oil concentrations.

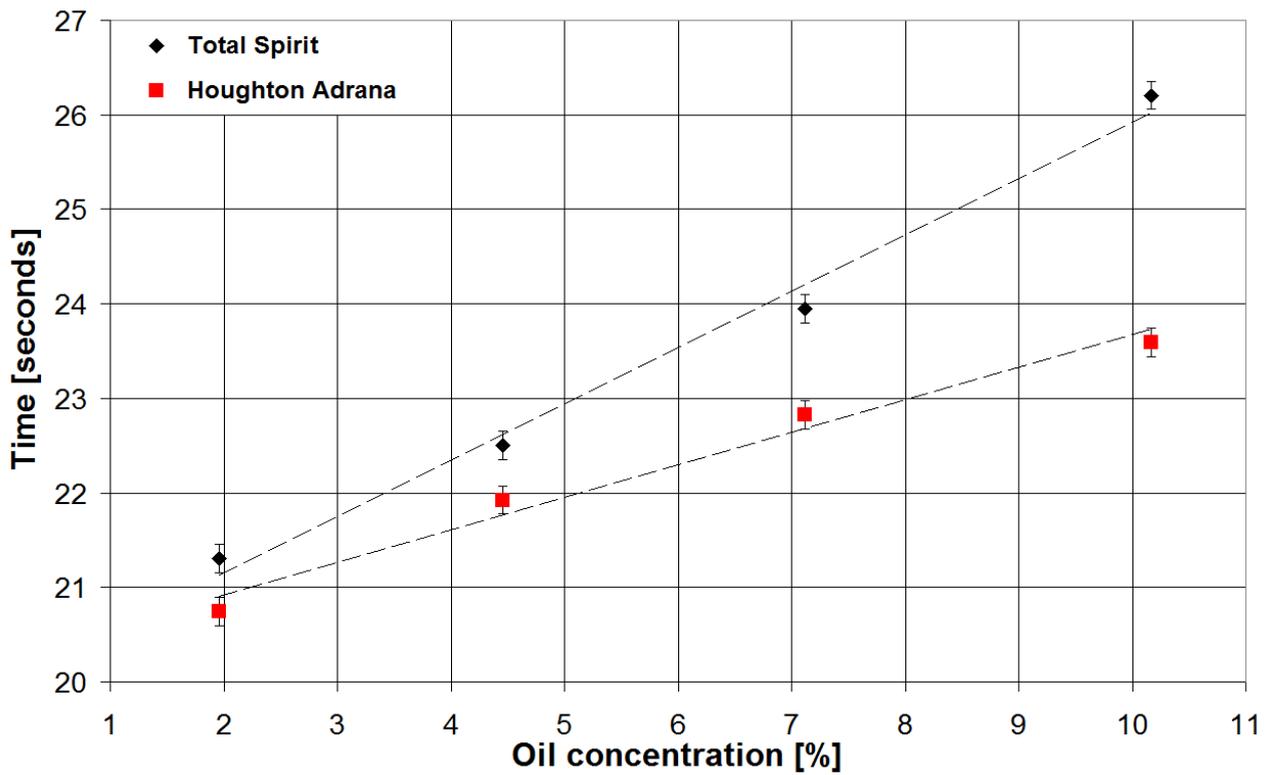

**Fig. 6**. Measured $t_{fall}$ vs $C_{oil}$ for two different products ("Spirit MS 8200" by Total and "Adrana D 208" by Houghton). $\vartheta = 10°$ and T = 25°C.

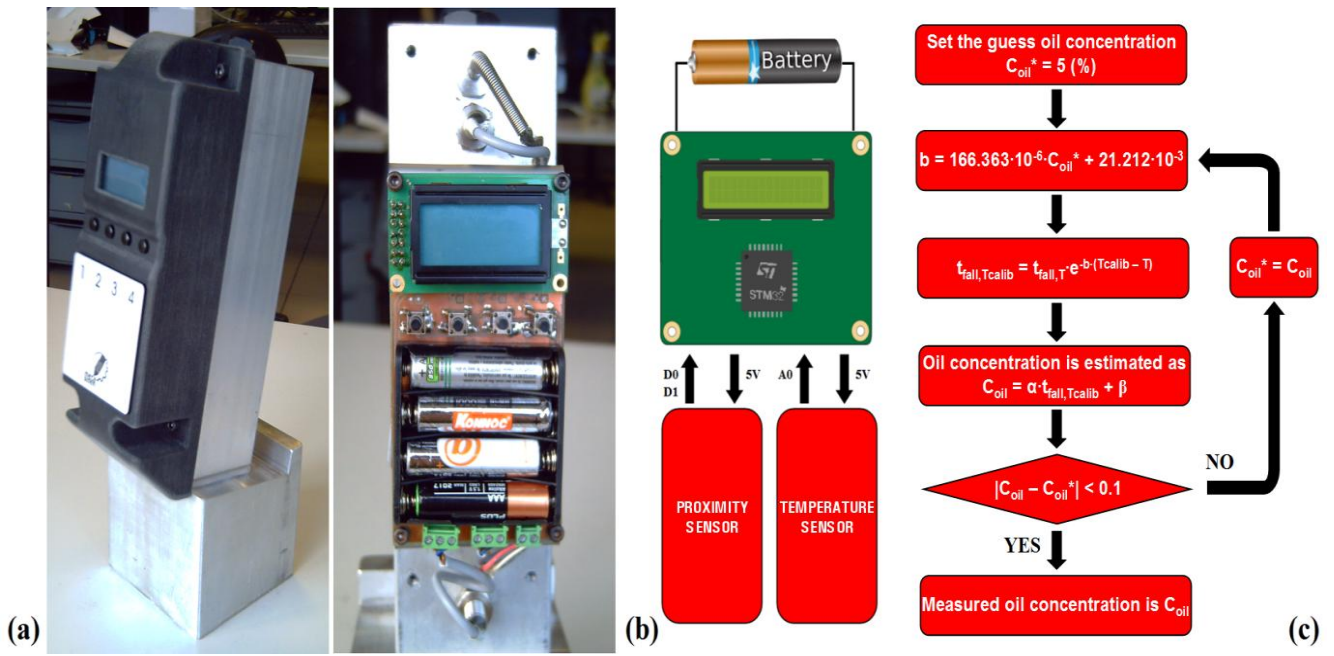

**Fig. 7** (a) Pictures and (b) schematic of the portable electronic system developed in this work. (c) Iterative procedure used to calculate the sample oil concentration.

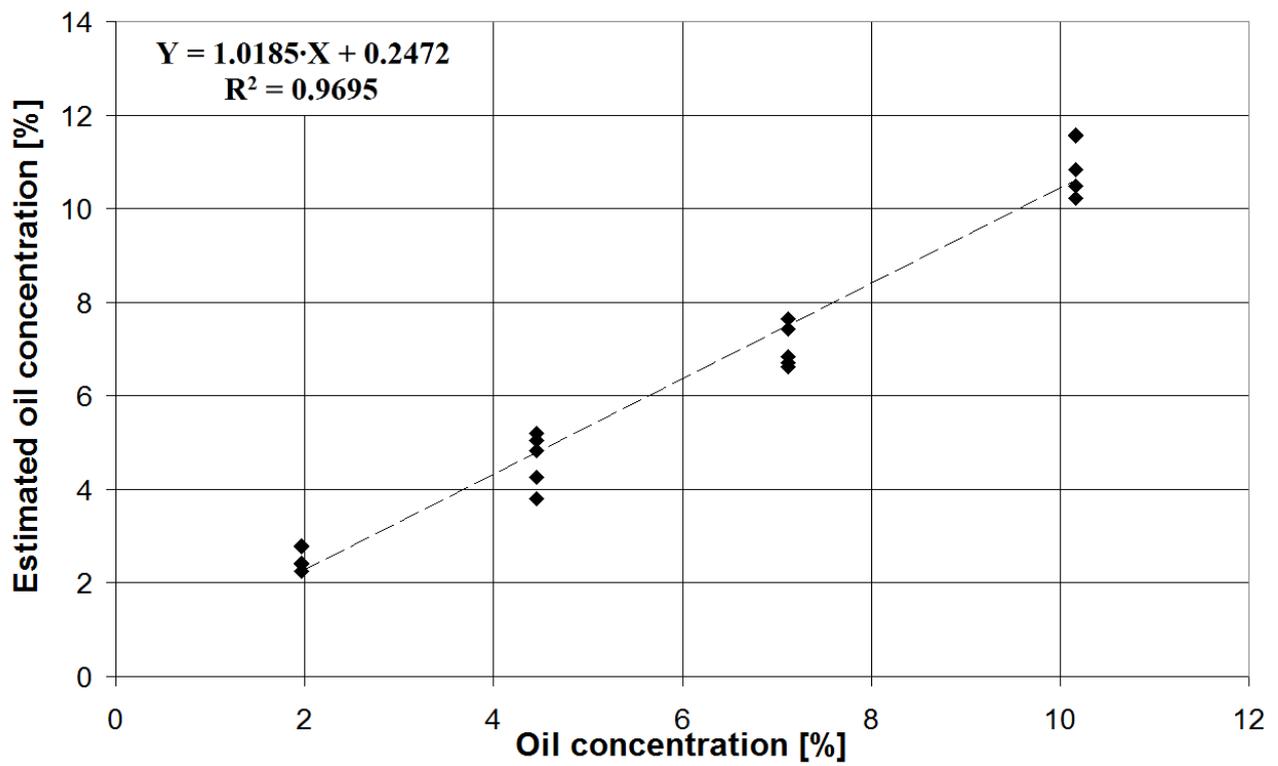

**Fig. 8** Scatter plot of estimated vs. real oil concentrations obtained with the portable system of this work.

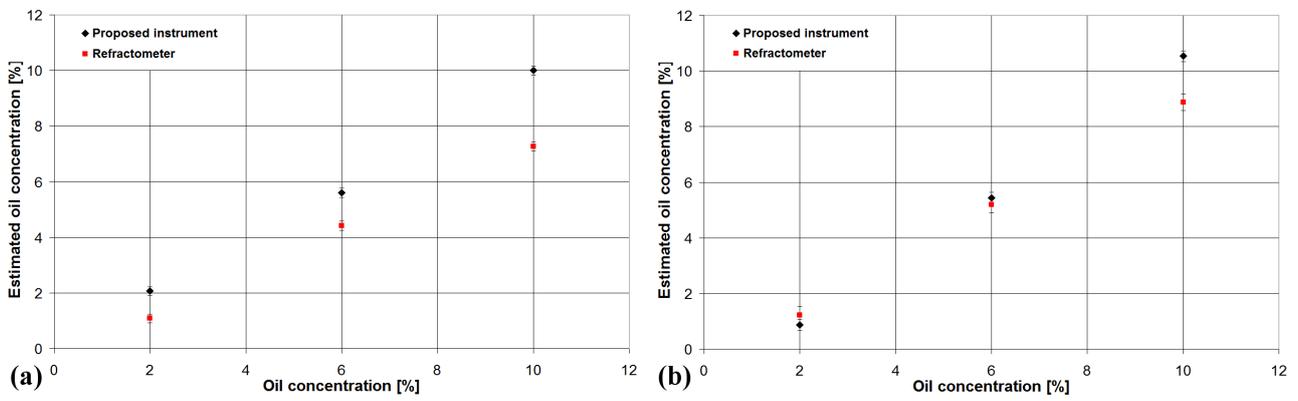

**Fig. 9** Estimated value of $C_{oil}$ vs. real oil concentration for the proposed instrument and a refractometer in the case of fresh samples prepared with (a) "Spirit MS 8200" by Total and (b) "Adrana D 208" by Houghton.